\documentclass[aps,prb,twocolumn,groupedaddress,nofootinbib]{revtex4}
\usepackage{graphicx}
\usepackage{hyperref}
\usepackage[english]{babel}
\usepackage{makeidx}
\usepackage[usenames]{color}
\definecolor{Red}{rgb}{1,0.00,0.00}

\begin{document}

\title{A Dipole Polarizable Potential for Reduced and Doped CeO$_2$ from First-Principles.}

\author {Mario Burbano $^1$} 
\author{Dario Marrocchelli$^2$} \email{dmarrocc@mit.edu}
\author{Bilge Yildiz $^2$} 
\author{Harry L Tuller $^3$} 
\author{Stefan T Norberg $^4$} 
\author{Stephen Hull $^5$}  
\author{Paul A Madden$^6$} 
\author{Graeme W. Watson$^1$} \email{watsong@tcd.ie}

\address{$^1$ School of Chemistry and CRANN, Trinity College Dublin, Dublin 2, Ireland}
\address{$^2$ Department of Nuclear Science and Engineering, Massachusetts Institute of Technology}
\address{$^3$ Department of Materials Science and Engineering, Massachusetts Institute of Technology}
\address{$^4$ Department of Chemical and Biological Engineering, Chalmers University of Technology}
\address{$^5$ The ISIS Facility, Rutherford Appleton Laboratory}
\address{$^6$ Department of Materials, University of Oxford, Parks Road, Oxford OX1 3PH, United Kingdom}


\begin{abstract}
In this paper we present the parameterization of a new interionic potential for stoichiometric, reduced and doped CeO$_2$. We use a dipole-polarizable potential (DIPPIM) and optimize its parameters by fitting them to a series of DFT calculations. The resulting potential was tested by calculating a series of fundamental properties for CeO$_2$ and by comparing them to experimental values. The agreement for all the calculated properties (thermal and chemical expansion coefficients, lattice parameters, oxygen migration energies, local crystalline structure and elastic constants) is within 10-15\% of the experimental one, an accuracy comparable to that of ab initio calculations. This result suggests the use of this new potential for reliably predicting atomic-scale properties of CeO$_2$ in problems where ab initio calculations are not feasible due to their size-limitations.
\end{abstract}

\newpage

\maketitle

\section{Introduction}
\label{Introduction}

Cerium dioxide, CeO$_2$ or ceria, is an important material which has found applications in several technologically relevant areas such as catalysis \cite{trovarelli} and Solid Oxide Fuel Cells (SOFCs) \cite{steele2001, jacobson2010}. 
In catalysis, it plays an important role thanks to its oxygen storage capability, due to the ready oxidation state change from Ce$^{4+}$ to Ce$^{3+}$ upon reduction and the reverse upon oxidation \cite{trovarelli}. These properties are made use of in Three-Way Catalysts (TWC), where the stored oxygen aids in the oxidation of CO to CO$_2$ under reducing conditions while, under fuel-lean conditions, the reduction of NO to N$_2$ is assisted by the uptake of oxygen by ceria. Doping ceria with aliovalent cations, such as Gd, Y or La, leads to high ionic conductivity in the intermediate temperature range (500 -- 800$^{\circ}$ C), thus raising the prospects of ceria-based electrolytes for application in SOFCs \cite{etsell1970, mogensen2000}.
\newline

Over the past 5 years, significant progress has been made in the description of this material by means of \textit{ab initio} computer simulations \cite{fabris2005, nolan2005a, nolan2005b, nolan2006, chen2007, fronzi2009, ganduglia2009}, using Density Functional Theory (DFT). In particular, the use of the DFT+U approach, where the U parameter provides an improved description of the strongly correlated cerium 4$f$ states in partially reduced ceria, has led to a much improved understanding of the electronic and structural properties of this material. Unfortunately, DFT calculations are still severely limited by system size and the time-scales that can be studied; this high computational cost usually limits this approach to static calculations only. For this reason, reliable interatomic potentials which allow the study of thousands of atoms on the nanosecond scale are desirable. This is particularly true for the study of the ionic conductivity of ceria. Indeed, the role of grain boundaries in the formation of space charge regions, or the vacancy and/or cation ordering tendencies, which are responsible for the drop in conductivity after a critical vacancy concentration (around 3-4 \%), are long-range in nature and necessitate large simulation boxes.
\newline

In a recent paper, Xu \textit{et al.} \cite{xu2010} compared six different interatomic potentials for ceria available in the literature \cite{grimes1997,vyas1998,inaba1999,gotte2004, gotte2007, butler2008} and tested their accuracy by reproducing a series of experimental data (lattice constants, thermal expansion, chemical expansion, dielectric properties, oxygen migration energy and mechanical properties). Two main limitations were found. The first was that none of the reviewed potentials could reproduce all the fundamental properties under study, although some displayed higher accuracy than others. While all the potentials could reproduce the static properties, such as lattice parameters and elastic constants, they all failed at reproducing the thermal expansion coefficient, and, to a lesser degree, the oxygen migration energy, for pure CeO$_2$. Indeed, some potentials gave thermal expansion coefficients which were one order of magnitude smaller than the experimental one and also severely underestimated the oxygen migration energy. Thermal and chemical expansion properties of ceria are particularly important in the context of SOFCs given that differential expansion of the components has a detrimental effect on the long term durability of the fuel cells \cite{kossoy2009, kossoy2010}. A second problem evinced from the study by Xu \textit{et al.} was that not all the interatomic potentials have a complete set of parameters available for the study of both doped and reduced CeO$_2$. The potential by Inaba \textit{et al} \cite{inaba1999}, for instance, properly reproduces the thermal expansion coefficient and the elastic properties, but cannot be tested for chemical expansion because it does not have parameters for Ce$^{3+}$. \newline

\begin{figure}[htbp]
\begin{center}
\includegraphics[width=8cm]{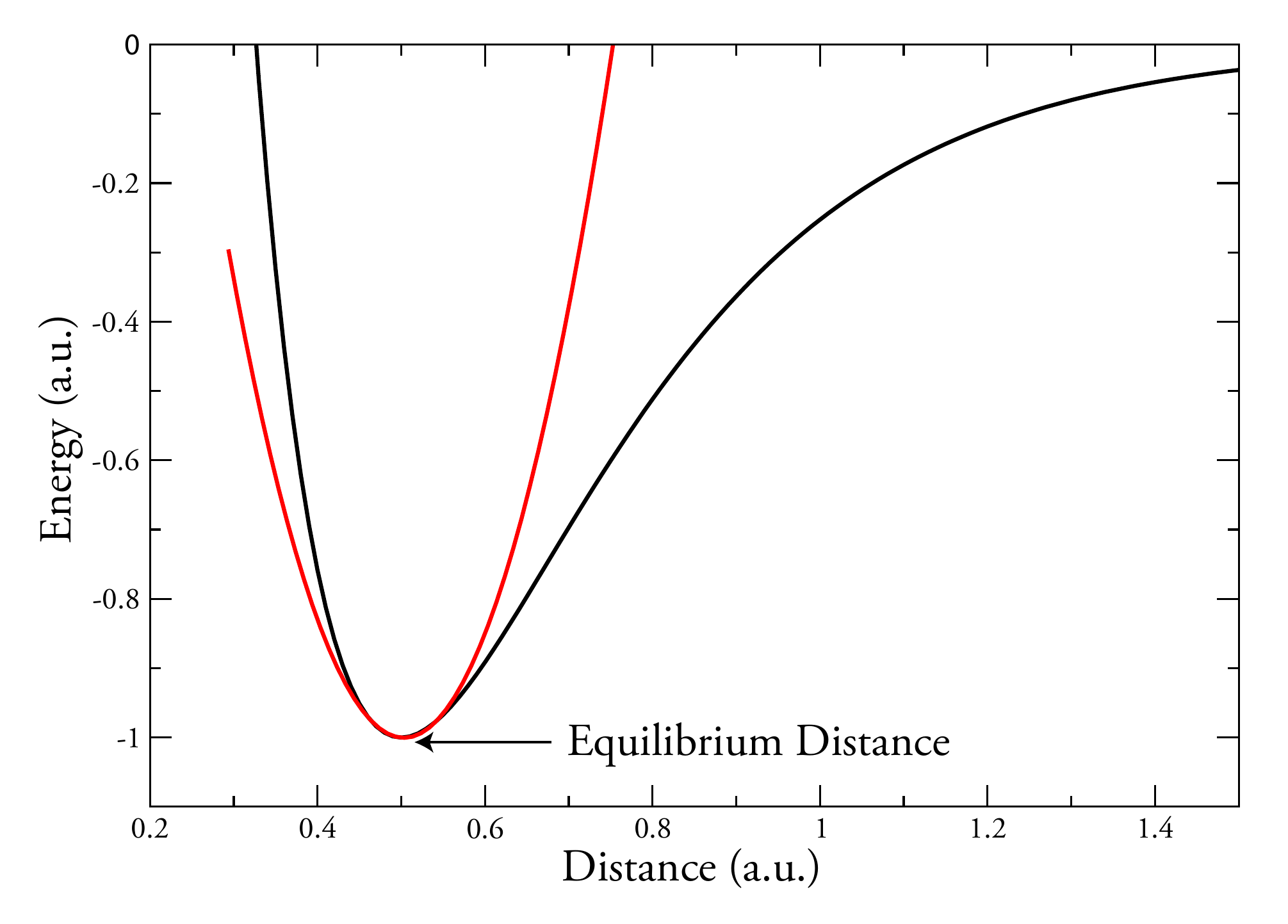}
\end{center}
\caption{The shape of a typical interatomic potential (black line) and of its harmonic approximation (red line).}
\label{Potential}
\end{figure}

The first limitation can be easily understood by looking at figure \ref{Potential}, where we show the shape of a typical interatomic potential. Such a potential is harmonic in the vicinity of the equilibrium position (red curve is a parabola fitted to the potential) but at distances away from the equilibrium position, it deviates from that shape and becomes anharmonic. It is this anharmonicity which is responsible for the thermal expansion observed in solid materials. A potential's failure to reproduce the experimental thermal expansion means that it is not properly parametrized and/or the potential's shape is not correct at distances greater than the equilibrium position. In particular the strong underestimation of the thermal expansion coefficient, as observed by Xu \textit{et al.}, indicates that the potential maintains a harmonic description of the system in regions where this approximation is not valid. The potential also deviates from the harmonic behavior at distances shorter than the equilibrium one, with the interatomic potential being more repulsive than its harmonic approximation. This feature plays an important role in the calculation of the oxygen migration barrier. When an oxygen ion hops from one site to another, it has to squeeze between neighboring cations, so that the average interatomic distance between the oxygen and these cations is much smaller than at equilibrium. If a potential's description of this interaction is predominantly harmonic, then the repulsion between these ions will be underestimated and, consequently, the migration energy barrier too will be underestimated, as observed by Xu \textit{et al.} \cite{xu2010}
\newline

It would seem from the above analysis that the limitations of the existing potentials are due to the fact that they are too harmonic. This might be due to the way they are parametrized, given that they are optimized by fitting their parameters to a small data set of experimental properties, usually elastic constants and lattice parameters. These are equilibrium or near-to-equlibrium properties, which allow the sampling of only a small region of the potential energy surface (PES) of the system, around the equilibrium distance. Little or no information is provided about the long- or short-distance behavior where the potential becomes anharmonic. In this paper we use a different methodology, which was used successfully for similar oxide systems \cite{wilson1996, wilson2004, jahn2007a, jahn2007b, marrocchelli2009a, marrocchelli2009b, marrocchelli2009c, norberg2009, marrocchelli2010, norberg2011, marrocchelli2011}, to parametrize interatomic potentials for stoichiometric, reduced and doped CeO$_2$. The key idea of this methodology is that the potential's parameters are fitted to a series of DFT calculations on high-temperature, distorted CeO$_2$ configurations. This allows sampling wider regions of the PES than those accessed by macroscopic equilibrium observables. We obtain parameters for stoichiometric, reduced and doped (La, Gd and Y) CeO$_2$ and test them against the experimental data. The agreement is quite good for all the studied properties, including thermal expansion and oxygen migration energy barriers. This methodology can be easily extended to other dopant cations in ceria or similar materials.

\section{Potential development}
\label{Potential development}
The interaction model used in this work is the same as that used in previous work on similar systems, such as GeO$_2$ \cite{marrocchelli2009d, marrocchelli2010}, doped ZrO$_2$ \cite{norberg2009, marrocchelli2009b, norberg2011, marrocchelli2011}, and Li$_2$O \cite{wilson2004}. The model \footnote{This model is implemented in an in-house molecular dynamics code called PIMAIM.}, known as DIPole Polarizable Ion Model (DIPPIM), includes a pair potential (a Buckingham term plus Coulombic interactions), together with an account of the polarization effects that result from the induction of dipoles on the ions. This model is conceptually similar to the shell model used by many authors \cite{grimes1997, vyas1998, inaba1999, gotte2004, gotte2007, butler2008}. Here we use formal ionic charges (O$^{2-}$, Ce$^{4+}$, Ce$^{3+}$, La$^{3+}$, Gd$^{3+}$) which should ensure better transferability. {A description of this model and the notation used for its parameters is reported in Appendix \ref{DIPPIM}}. The parameters for these potentials were obtained by matching the forces/dipoles obtained from DIPPIM to first-principles reference data \cite{madden2006a}. Such an approach has been applied successfully in the case of other oxide materials \cite{wilson1996, wilson2004, jahn2007a, jahn2007b, marrocchelli2009a, marrocchelli2009b, marrocchelli2009c, norberg2009, marrocchelli2010, norberg2011, marrocchelli2011}. In the following we give a brief description of the first-principles-based reference calculations and the force/dipole-matching procedure. \newline

First, a potential was parametrized for Y doped CeO$_2$, since this system will be the object of a detailed experimental and computational study by the authors of this paper \cite{norberg2011b} . This was done by performing DFT calculations, in the local density approximation (LDA), using the CPMD code \cite{CPMD}. For this system, the simple DFT approach is known to give the correct valence states (Ce$^{4+}$ and Y$^{3+}$), so that a DFT+U functional was not needed. Twelve 2 x 2 x 2 supercells with Ce$_{0.5}$Y$_{0.5}$O$_{1.75}$ compositions and a total of 88 atoms were constructed. Each model supercell was obtained from high temperature (2500\,K) MD simulations that were run for 50\,ps in order to reach structural equilibrium. The forces on each species were determined directly from each DFT calculation, and the dipoles were obtained from a Wannier analysis of the Kohn-Sham (KS) wave functions \cite{marzari1997}. 
Once the information about {\em both} forces and dipoles was gathered from the \textit{ab initio} calculations, the parameters in the interatomic potential were fitted to them. One problem with DFT calculations is the uncontrolled representation of the dispersion interaction. Although dispersion energies constitute only a small fraction of the total energy, they have a considerable influence on transition pressures and, in particular, on the material density and stress tensor. For this reason, the dispersion parameters were not included in the fits but were added afterwards, as discussed by Madden \textit{et al.} \cite{madden2006a}. The parameters from refs. \cite{jahn2007b, norberg2009} were used. The resulting parameters for the potential are reported in table  \ref{Table1}. The short-range parameters for the O-Y interaction are in line with those obtained in a previous study \cite{norberg2009} and the value for the O$^{2-}$ polarizability is close to that obtained from independent {\it ab-initio} calculations \cite{heaton2006b}. 
 \newline

A potential was parametrized for reduced CeO$_2$ as well. Similarly as above, first a short \textit{ab initio} MD simulation was performed on a 2 x 2 x 2 supercell with CeO$_{1.875}$ composition at high temperature. This configuration was used to calculate the forces acting on the ions and these, in turn, were used for the fitting procedure. These calculations were performed with the Vienna Ab-Initio Simulation Package (VASP) \cite{kresse1994, kresse1996, kresse1999} within the DFT+U framework. A value of U $=$ 7\,eV was chosen to ensure correct localization of the $f$ electrons. Although this is slightly higher than is generally used \cite{nolan2005a, nolan2005b}, it was found to be necessary to ensure localization in the non-equilibrium structures used in the potential fitting. We used the Generalized Gradient Approximation (GGA) with the Perdew-Wang 91 (PW91) exchange-correlation functional and an energy cut-off of 400 eV. The Ce$^{3+}$ cations were identified as those with spin 1, while the Ce$^{4+}$ ones have no spin. The Ce$^{3+}$ and Ce$^{4+}$ cations are then treated as two {\em different} cationic species in the potential parametrization and separate terms are obtained to describe the interactions between, for instance, Ce$^{4+}$ - O$^{2-}$ and Ce$^{3+}$ - O$^{2-}$. The parameters for the Ce$^{4+}$ - O$^{2-}$ and O$^{2-}$ - O$^{2-}$ were then fixed to the values obtained before for the Y doped CeO$_2$ system, so that these parameters are consistent and can be used all together. This procedure is equivalent to treating the Ce$^{3+}$ cations as a dopant species, in the same way as was done for Y. While an ionic model for reduced CeO$_2$ might not be justified \textit{a priori}, the picture arising from DFT+U calculations and experiments seems to confirm this model. The $f$ electrons are, indeed, found to be strongly localized on the Ce$^{3+}$ cations in agreement with an ionic picture.  Also, a recent study on Ce$^{3+}$/Ce$^{4+}$ ordering in ceria nanoparticles \cite{migani2009} showed that strikingly similar relative energy ordering of the isomers and atomic scale structural trends (e.g., cation--cation distances) are obtained in both the DFT and interionic-potential calculations, which, again, proves the validity of this approach. The resulting parameters for the Ce$^{3+}$ cations are reported in table \ref{Table1}. The potential parametrization was performed using two DFT codes because the CPMD package does not have an implementation of the DFT+U framework needed to describe reduced CeO$_{2}$, while the functionality required for the Wannier analysis was not available in VASP at the time.\newline

Finally, the same procedure can be repeated to include other cations. In this case we obtained parameters for La$^{3+}$ and Gd$^{3+}$. This was again done by performing a short MD simulation on Gd and La-doped CeO$_2$ at high temperature and using the final configuration to calculate forces to which the potential parameters can be fitted. Once again, we fixed the  parameters for the Ce$^{4+}$ -- O$^{2-}$ and O$^{2-}$ -- O$^{2-}$ to the values obtained before for the Y doped CeO$_2$ system. This ensures that these parameters are all consistent and allows the study of mixed systems -  such as, for instance, partially reduced, and Y and La doped CeO$_2$. The resulting parameters for the dopant cations are reported in table \ref{Table1}. This procedure can be repeated for as many dopant cations as desired. 


\begin{table*}[htbp]
\caption{\label{Table1} Parameters of the DIPPIM potential. All values are in atomic units. A description of the parameters is reported in \ref{DIPPIM}. For the polarizability part of the potential we report only those parameters with b not equal to zero.}
\begin{center}
\begin{tabular}{c c c c c c c}
\noalign{\smallskip}\hline\noalign{\smallskip}
         	      & O-O 		& Y-O 	& Ce$^{4+}$-O & Ce$^{3+}$-O  & Gd-O	& La-O \\
\hline
\hline
$A^{ij}$ &  55.3     	& 111.1 	& 105.9 		& 218.7 		& 236.9	& 98.6   \\
$a^{ij}$ &  6.78     	& 1.377 	& 1.269 		& 1.473 		& 1.566	& 1.257 \\
$B^{ij}$ & 50000 	& 50000 	& 50000 		& 50000  		& 50000	& 50000\\
$b^{ij}$ & 0.85   	& 1.35   	& 1.4   		& 1.35    		& 1.35	&1.35     \\
\\
$C_6^{ij}$ & 53  	& 12  	& 12  		& 12 			& 12		& 12\\
$C_8^{ij}$ & 1023 	& 240 	& 240 		& 240 		& 240	& 240 \\
$b_{disp}^{ij}$ & 1.0  & 1.5  	& 1.5 		& 1.5 		& 1.5		& 1.4\\
\\
$\alpha_{O^{2-}}$   & 14.9   	& 	& 	&  		&		&\\
$\alpha_{Y^{3+}}$   & 2.60   	& 	& 	&   		&		&\\
$\alpha_{Ce^{4+}}$  & 5.0   	& 	& 	&  		&		&\\
$\alpha_{Ce^{3+}}$  & 11.2   	& 	& 	&  		&		&\\
$\alpha_{Gd^{3+}}$  & 6.8   	& 	& 	&  		&		&\\
$\alpha_{La^{3+}}$  & 10.8   	& 	& 	&  		&		&\\
\\
\hline
\hline
\\
$b_D^{O-O}$ 		& 1.73     	&  $b_D^{O-Ce^{4+}}$ 	& 1.76	&  $b_D^{Ce^{4+}-O}$ 	&1.76     \\
$c_D^{O-O}$ 		& 0.45 	&  $b_D^{O-Ce^{4+}}$  	& 1.93	&  $b_D^{Ce^{4+}-O}$ 	&-.47    \\
$b_D^{O-Ce^{3+}}$ & 1.82       &  $b_D^{Ce^{3+}-O}$ 	& 1.82	&  $b_D^{O-Y^{3+}}$ 	&1.67\\
$c_D^{O-Ce^{3+}}$ & 2.92       &  $b_D^{Ce^{3+}-O}$  	& -2.50	&  $b_D^{O-Y^{3+}}$ 	&1.62 \\

$b_D^{Y^{3+}-Y^{3+}}$ & 1.67       &  $b_D^{Y^{3+}-O}$ 	& 0.59	&  $b_D^{Y^{3+}-Ce^{4+}}$ 	&0.75\\
$c_D^{Y^{3+}-Y^{3+}}$ & 0.90       &  $b_D^{Y^{3+}-O}$  	& -.33	&  $b_D^{Y^{3+}-Ce^{4+}}$ 	&-.45 \\

$b_D^{O-Y^{3+}}$ & 1.93       &  $b_D^{Gd^{3+}-O}$ 	& 1.94	&  $b_D^{O-La^{3+}}$ 	& 1.69\\
$c_D^{O-Y^{3+}}$ & 2.81       &  $b_D^{Gd^{3+}-O}$  	& -.44	&  $b_D^{O-La^{3+}}$ 	& 2.02 \\
$b_D^{La^{3+}-O}$ & 1.69       & & & & \\
$c_D^{La^{3+}-O}$ & -1.21       &  & & &  \\

\\
\end{tabular}
\end{center}
\end{table*}

\section{Results}
\label{Results}

The quality of the potentials can be assessed by comparing model predictions to experimental data, since no experimental data was used in the optimization of the model parameters. In this section we present our predictions, using the herein developed interatomic potentials, on lattice parameters, local crystalline structure, thermal and chemical expansion, oxygen migration energies and elastic constants, and their comparison with the experimental values. The way these quantities are calculated is described in each section.

\subsection{Thermal expansion}

\begin{figure}[htbp]
\begin{center}
\includegraphics[width=8cm]{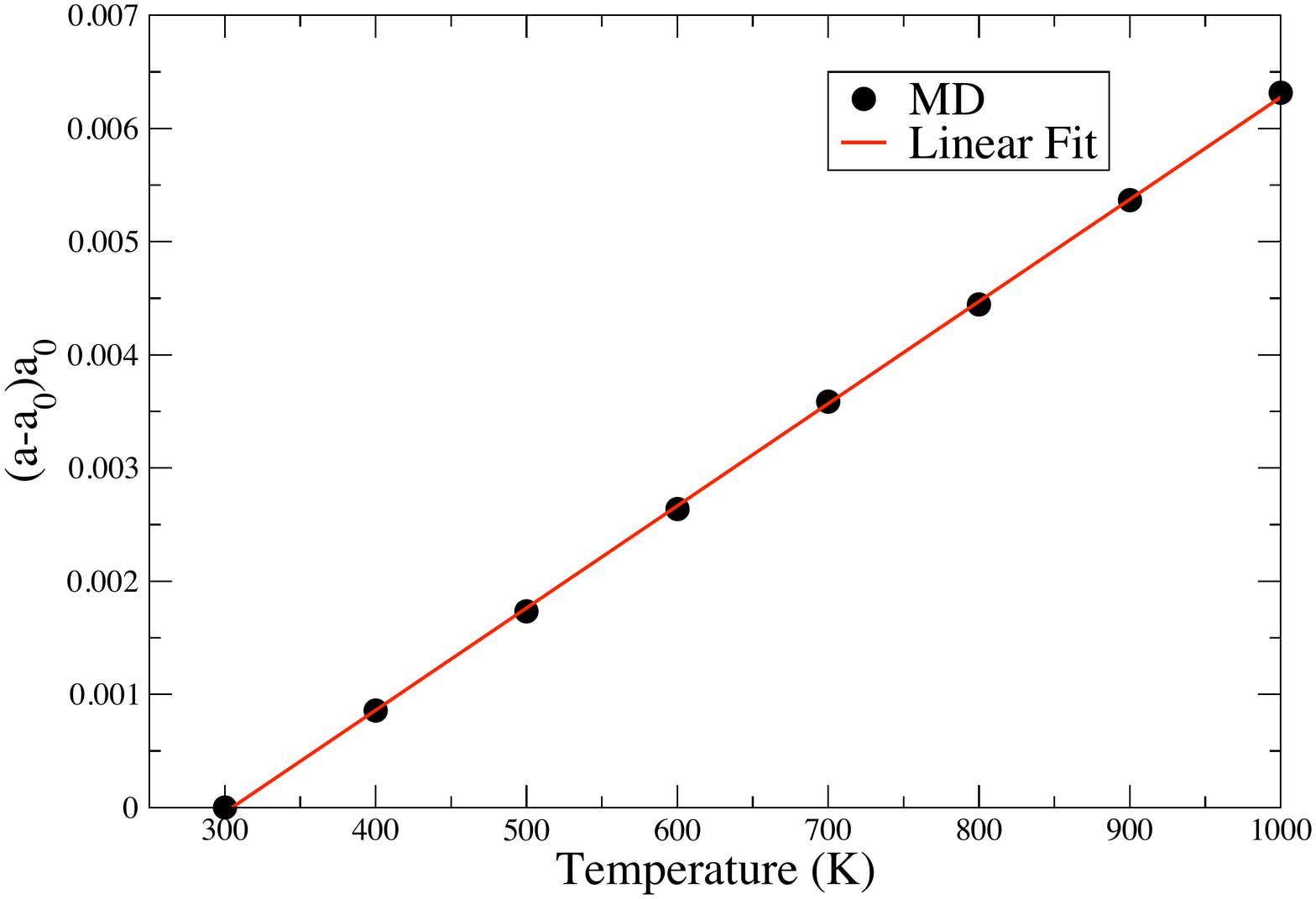}
\end{center}
\caption{Simulated lattice expansion (points) versus temperature for stoichiometric CeO$_2$. The red line represents a linear fit of the MD data. The thermal expansion, $\alpha$, extrapolated from the fit is 9.0 x $10^{-6}$ K$^{-1}$.}
\label{ThermalExpansion}
\end{figure}

Xu \textit{et al.} \cite{xu2010} find thermal expansion to be one of the most difficult properties to model in pure ceria. Figure \ref{ThermalExpansion} shows the calculated expansion of the lattice parameter, in the 300 - 1000 K temperature range. This is defined as (a-a$_0$)/a$_0$, where a is the lattice parameter for a certain temperature and a$_0$ is the lattice parameter at 300 K. This was obtained by performing molecular dynamics simulations with 4 x 4 x 4 supercells in an NPT ensemble, at the required temperatures. We used barostats and thermostats as described by Martyna \textit{et al.} \cite{martyna1992a, martyna1994a} and we set the external pressure to zero. The lattice parameters were averaged over a 0.1 ns long simulation and used to calculate the percentage expansion. The corresponding thermal expansion coefficient is extracted by fitting a straight line to our data (see figure \ref{ThermalExpansion}). The obtained value is $\alpha$ = 9.0 x $10^{-6}$ K$^{-1}$ which is within 20 \% of the experimental values (see table \ref{Table2}). This is a substantial improvement compared to the Grimes \cite{grimes1997}, Gotte 2004 \cite{gotte2004} and Gotte 2007 \cite{gotte2007} potentials which gave a thermal expansion coefficient of 1.27 x $10^{-6}$ , 6.65 x $10^{-6}$ , 7.31 x $10^{-6}$  K$^{-1}$, respectively. We remind the reader that no empirical data was used in the parameterization of this potential, so that this potential was not manually optimized in order to reproduce the thermal expansion coefficient.

\begin{center}
\begin{table}[h]
\begin{tabular}{l c}
\hline
Thermal expansion	coefficient	 ( $10^{-6}$ K$^{-1}$)	& Reference 	 \\
 							&			 \\
\hline
\hline
9.0 			    			&  This work		  \\
\hline
10.7 						& Ref. \cite{hisashige2006}		\\
\hline	 
11.1	 					&  Ref. \cite{sameshima2002}		\\
\hline		 
11.6	 					&  Ref. \cite{sims1976}		\\
\hline	
\end{tabular}
\caption{Comparison between the experimental and simulated thermal expansion coefficients.} \label{Table2}
\end{table}
\end{center}

\subsection{Elastic properties}
Elastic constants and the bulk modulus were calculated and compared to experimental values. The three independent elastic constants were obtained by straining an optimized simulation cell in different directions by a small amount (typically a fraction of a percent) and the resulting stress tensor is recorded after relaxation of the atomic positions. After repeating this procedure for several magnitudes of positive or negative strain, the linear relationship between strain and stress is used to obtain the elastic constants at 0 K. The bulk modulus was extracted from a volume versus pressure curve. \newline

Table \ref{Table4} reports the three elastic constants and the bulk modulus. These are within 15\% the reported experimental values, with the exception of C$_{11}$, whose value is overestimated by 32\%. We believe this agreement to be quite good, especially considering that the experimental data for these properties are quite scattered. \newline

\begin{table}[h]
\begin{tabular}{c c c}
\hline
Property  (GPa)	& MD & Experimental \cite{nakajima1994, clausen1987}	 \\
 						&	&		 \\
\hline
\hline
C$_{11}$	    		&  552	& 386 - 450	\\
\hline	 
C$_{12}$ 			& 137 	& 105 -- 124	\\
\hline	
C$_{44}$ 			& 66  	& 60 -- 73	\\
\hline	
B 				& 275	& 204 -- 236 	\\
\hline				
\end{tabular}
\caption{Experimental and simulated elastic constants and bulk modulus for stoichiometric CeO$_{2}$.} \label{Table4}
\end{table}

\begin{figure}[htbp]
\begin{center}
\includegraphics[width=8cm]{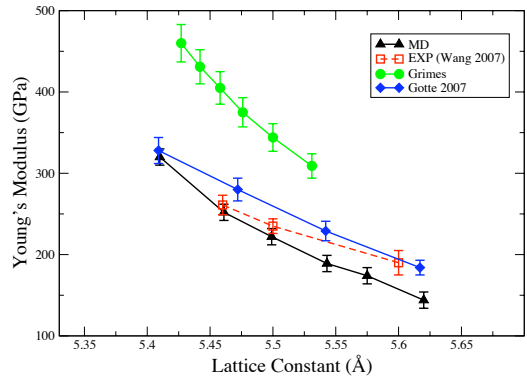}
\end{center}
\caption{Young's modulus versus the lattice parameter in CeO$_{2-x}$. Red symbols are the experimental data from Wang \textit{et al.} \cite{wang2007}, green and blue symbols are the simulated data obtained with the Grimes \cite{grimes1997} and Gotte 2007 \cite{gotte2007} potentials, while the black ones are from our own potential.}
\label{YoungModulus}
\end{figure}

One of the main motivations of the work of Xu \textit{et al.} \cite{xu2010} was to find a reliable potential to describe - and perhaps explain - the observed elastic softening of ceria with decreasing oxygen partial pressure ($P_{\rm{O_2}}$).  {We therefore calculated the Young's modulus by using the following formula \cite{Kanchana2006},
$$
E = \frac{9BG}{3B+G},
$$
where B is the bulk modulus and G the shear modulus, which can be obtained from the elastic constants \cite{Kanchana2006}.} Figure \ref{YoungModulus} reports the Young's modulus for reduced ceria as a function of the lattice constant, together with the experimental data from the work of Wang  \textit{et al.} \cite{wang2007, xu2010} and the ones obtained with the Grimes and Gotte 2007 potentials,  {as reported by Xu {\it et al.}} \cite{xu2010}. The scale of the plot is the same as in Fig. 3 in ref \cite{xu2010} to facilitate a comparison. Our interatomic potential shows the expected elastic softening as a function of the lattice constant (or analogously, as a function of non-stoichiometry).  {When compared to the experimental data, it seems that our potential slightly overestimates this softening, although most of the values obtained with our simulations are within experimental error}. The level of accuracy is indeed better than the Grimes potential and comparable with the Gotte 2007 potential.  {Finally, preliminary calculations on Ce$_{1-x}$$M_{x}$O$_{2-x/2}$, with M = Y$^{3+}$, Gd$^{3+}$ and La$^{3+}$,  indicate that this system shows a similar softening as a function of the dopant concentration, $x$.}

\subsection{Structural properties of reduced CeO$_2$}
The local crystal structure of reduced ceria has been recently studied by neutron diffraction \cite{hull2009}. From that set of data, total radial distribution functions were extracted for different values of $x$ in CeO$_{2-x}$. Total radial distribution functions can be expressed in terms of the individual partial radial distribution functions, $g_{ij}(r)$, weighted by the concentrations of the two species, $c_i$ and $c_j$, and their coherent bound neutron scattering lengths, $b_i$ and $b_j$, so that

$$ G(r)=\sum _{i,j=1}^{n} c_i c_j b_i b_j g_{ij}(r) / \sum_{i=1}^n (b_i c_i)^2, $$

where n is the number of ionic species. The partial radial distribution functions $g_{ij}(r)$ are given by:

$$ g_{ij}=\frac{1}{4\pi r^2 \Delta r} \frac{n_{ij}(r)}{\rho_j},$$

with $n_{ij}(r)$ equal to the number of atoms of type $j$ located at a distance between $r$ and $r + \Delta r$ from an atom of type $i$ and $\rho_j$ is the number density of atoms of type $j$, given by $\rho_j = c_j \rho_0$. These partial radial distribution functions can be easily calculated from the simulation output. \newline

\begin{figure}[htbp]
\begin{center}
\includegraphics[width=8cm]{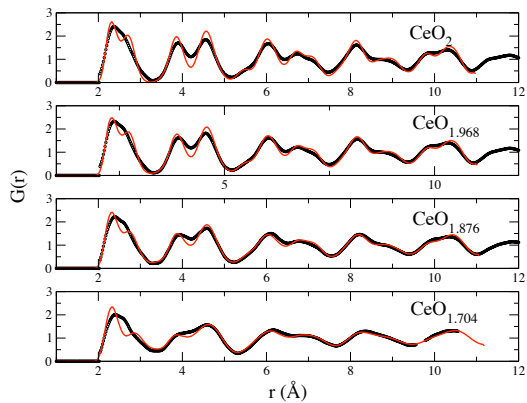}
\end{center}
\caption{Radial distribution functions, $G(r)$, for different values of $x$ in CeO$_{2-x}$, at 1273 K. Red lines and black empty dots correspond to the MD and experimental \cite{hull2009} $G(r)s$ respectively.}
\label{RDFs}
\end{figure}

In figure \ref{RDFs}, we report the calculated total radial distribution functions, $G(r)$, for different non-stoichiometries and we compare them with those extracted from the neutron diffraction data. The agreement is good for all the studied compositions. 
A visual analysis of the oxygen-oxygen radial distribution functions (not shown) shows an increased broadening as a function of $x$, which is indicative of an increased disorder within the anion sublattice, in agreement with experiments \cite{hull2009}.

\subsection{Chemical expansion}
\begin{figure}[htbp]
\begin{center}
\includegraphics[width=8cm]{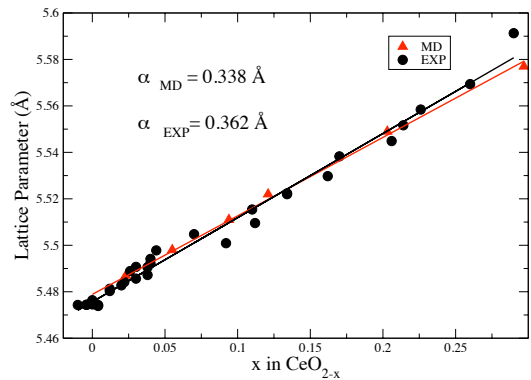}
\end{center}
\caption{MD lattice parameter (red triangles) as a function of the nonstoichiometry, $x$, in CeO$_{2-x}$, with the corresponding linear fit (red line). The simulations were performed at 1273 K, the same temperature as the reported experimental data \cite{hull2009} (black circles). The black line is a linear fit to the experimental data.}
\label{LattVSx}
\end{figure}

The conditions found at the anode side of SOFCs lead to the reduction of Ce$^{4+}$ to Ce$^{3+}$ with a subsequent change in the lattice parameter. This chemical expansion affects the performance of the electrolyte as it creates a strain in the cell and can eventually cause fracture. For this reason, in this section, we test the ability of our potential to describe this behavior accurately. In figure \ref{LattVSx} we report the calculated lattice parameter as a function of the oxygen non-stoichiometry in CeO$_2$ at 1273 K and compare this with the neutron data from Hull \textit{et al.} \cite{hull2009} at the same temperature. The agreement is excellent and the simulated chemical expansion coefficient, 0.338\,\AA, is within 7 \% of the experimentally determined chemical expansion coefficient, 0.362\,\AA. Such a good agreement is encouraging and also indirectly confirms that our ''ionic'' approach, in which we see Ce$^{3+}$ as a different cation species, carries the correct physics. \newline

In table \ref{Table5} we report the calculated lattice parameters for Ce$_{0.8}$$M_{0.2}$O$_{1.9}$ at room temperature, where $M$ $=$ Gd$^{3+}$, La$^{3+}$ and compare them with the experimental values. The agreement is within 1\% and the trend of increasing lattice parameter with increasing cation radius is properly reproduced.

\begin{table}[h]
\begin{tabular}{c  c  c}
\hline
Lattice parameter	& MD (\AA) & Experimental (\AA) \cite{sameshima2000}	 \\
 						&	&		 \\
\hline
\hline
Gd$^{3+}$	    		&  5.426	& 5.423	\\
\hline	 
La$^{3+}$ 			&  5.494 	& 5.476	\\
\hline				
\end{tabular}
\caption{Comparison between experimental \cite{sameshima2000} and simulated lattice parameters for Ce$_{0.8}$$M_{0.2}$O$_{1.9}$ at room temperature ($M$ $=$ Gd$^{3+}$, La$^{3+}$).} \label{Table5}
\end{table}

\subsection{Oxygen migration energies}
\begin{figure}[htbp]
\begin{center}
\includegraphics[width=8cm]{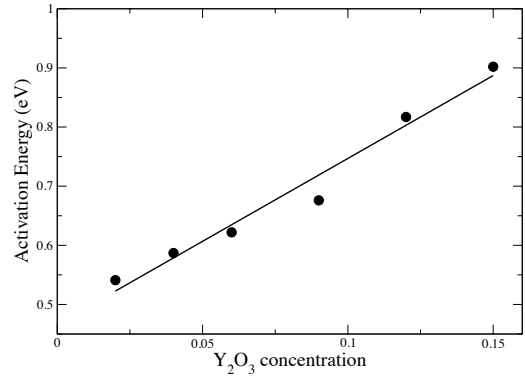}
\end{center}
\caption{Activation energy (black points) in Ce$_{1-x}$Y$_x$O$_{2-x/2}$ versus Y concentration, $x$, with linear fit (black line). These energies were extracted from an Arrhenius plot of the MD diffusion coefficients at high temperatures.}
\label{MigrationEnergyVSx}
\end{figure}

 {In pure ceria, the activation energy ($E_{a}$) which determines the ionic conductivity at a given temperature is composed of the vacancy formation enerergy ($E_{f}$) and the oxygen migration energy ($E_{m}$). In this case, the low concentration of oxygen vacancies results in a low ionic conductivity. Aliovalent doping generates vacancies through charge compensation so that the activation energy then becomes the sum of the the migration energy plus the association (binding) energy ($E_{b}$) between the dopant cations and the vacancies. \cite{arachi1999, politova2004, ruehrup2006, andersson2006}.} Molecular dynamics simulations on the Ce$_{1-x}$Y$_x$O$_{2-x/2}$ system were performed for different values of $x$ in order to evaluate the diffusion coefficient and the relative oxygen migration energies. These simulations were performed on 4 x 4 x 4 supercells at a various Y$_2$O$_3$ concentrations. Each concentration was initially equilibrated at a temperature of 2073\,K for 120 ps, using a timestep of 1 fs. The diffusion coefficients were then extracted from the oxide ion mean squared displacement for temperatures of 1073\,K and above, as explained by Norberg \textit{et al.} \cite{norberg2009}. The diffusion coefficients were then plotted as a function of the temperature and the oxygen migration activation energy was extracted. { {In principle, this activation energy contains both the migration energy and the association energy between the vacancy and the dopant cation. However, in doped-ceria conductors at high temperatures, which is the situation of interest here, isolated vacancies migrate freely so that the activation energy is equal to the migration energy only. 
\cite{arachi1999, politova2004, ruehrup2006}.}} Figure \ref{MigrationEnergyVSx} therefore shows the predicted activation energy (oxygen migration energy) as a function of the Y concentration, $x$. The observed behavior of an increasing migration energy as a function of Y concentration is in excellent agreement with the experimental findings of Tian \textit{et al.} \cite{tian2000}. \newline

If the migration energy in figure \ref{MigrationEnergyVSx} is extrapolated to x $=$ 0, a value of approximately 0.47\,eV is obtained. This value corresponds to the oxygen migration energy  {at infinite dilution, i.e. when the oxygen migration energy is unaffected by the dopant cations,} and compares well with the experimental results and with previous DFT calculations, as shown in table \ref{Table3}. 
In conclusion, this potential can successfully reproduce the ionic conduction properties of this material. A more detailed analysis of the ionic conduction mechanisms in doped CeO$_2$ and the factors that affect it will be the subject of a subsequent paper \cite{norberg2011b}.

\begin{table}[h]
\begin{tabular}{c c}
\hline
Oxygen migration energy	(eV)	& Reference 	 \\
 						&			 \\
\hline
\hline
0.47 			    			&  This work 		  \\
\hline	 
0.52						&  Ref. \cite{steele1971}		\\
\hline	 
0.40						&  Ref. \cite{tuller1977}	\\
\hline	
0.47						&  Ref. \cite{dholabhai2010}	\\
\hline	
\end{tabular}
\caption{Comparison between experimental and simulated oxygen migration energies for CeO$_{2-x}$, for $x\rightarrow 0$.} \label{Table3}
\end{table}

\section{Conclusions}
\label{Conclusions}
Driven by the recent work of Xu \textit{et al.}, which showed that none of the potentials for CeO$_2$ reported in the literature could reproduce all the fundamental properties of this system, in this paper we have presented the parameterization and the accuracy of a new interionic potential for stoichiometric, reduced and doped CeO$_2$. We used a dipole-polarizable (DIPPIM) potential and optimized its parameters by fitting them to a series of DFT calculations. The resulting potential was tested by calculating a series of fundamental properties for CeO$_2$ and by comparing them to experimental values. The agreement for all these properties (thermal and chemical expansion coefficients, lattice parameters, oxygen migration energies, local crystalline structure and elastic constants) is very good, with the calculated values being generally within 10-15\% of the experimental ones. We note that such accuracy is comparable to that of DFT calculations, but the computational cost is reduced significantly. \newline

These potentials can be used to predict and elucidate the atomic-scale properties of CeO$_2$ in situations where DFT calculations are not practical due to their size limitations. With such potentials, nanosecond long simulations on 1000s of atoms can be performed and these can be used to understand the structural, chemical, mechanical and conducting properties of this material, as previously done for similar systems \cite{wilson1996, wilson2004, jahn2007a, jahn2007b, marrocchelli2009a, marrocchelli2009b, marrocchelli2009c, norberg2009, marrocchelli2010, norberg2011, marrocchelli2011}.

\section*{Acknowledgements}
DM would like to thank Sean Bishop and Yener Kuru (MIT) for useful discussions and pointing out the limitations of the existing interionic potentials and Sandro Jahn (GFZ German Research Centre for Geosciences) for help in calculating the elastic constants. BY acknowledges Schlumberger-Doll Research Center for her support on ceria in reducing environments. DM and BY acknowledge the National Science Foundation for computational support through the TeraGrid Advanced Support Program, with a Start-up allocation (TG-DMR100098) and a Research allocation (TG-DMR110004). STN wishes to thank Vetenskapsr\aa det (Swedish Research Council) for financial support. HLT acknowledges Basic Energy Science, Department of Energy (DESC0002633), for his support on chemo-mechanics of ceria.  GWW acknowledges the Science Foundation Ireland research frontiers programme (grant numbers 08/RFP/MTR1044 and 09/RFP/MTR2274) and Trinity Center for High-Performance Computing (TCHPC). 

\appendix
\section{The DIPPIM model} \label{DIPPIM}
 {
In this section we report a brief description of the potential model employed in this work. The reader is referred to \cite{ norberg2009, madden2006a, castiglione1999a} and references therein for further reading. The interatomic potential is constructed from four components: charge-charge, dispersion, overlap repulsion and polarization. The first three components are purely pairwise additive: 
\begin{equation}
V^{\rm qq}=\sum_{i\leq j}\frac{q_i q_j}{r_{ij}}
\end{equation}
where $q_i$ is the {\it formal} charge on ion $i$. The dispersion  interactions include dipole-dipole and dipole-quadrupole terms
\begin{equation}
V^{\rm disp}=-\sum_{i\leq j }[\frac {f_6^{ij} (r^{ij})  C_6^{ij}}{r_{ij}^6
}+\frac{f_8^{ij} (r^{ij} ) C_8^{ij}}{r_{ij}^8 }].
\end{equation}
Here $C_6^{ij}$ and $C_8^{ij}$ are the dipole-dipole and  dipole-quadrupole dispersion coefficients, respectively, and the $f_n^{ij}$ are the Tang-Tonnies dispersion damping function, which describe short-range corrections to the asymptotic dispersion term. The short range repulsive terms are approximately exponential in the region of physical interionic separations. The full expression used here for the short range repulsion is:
\begin{eqnarray}
  V^{\rm rep}= \sum_{i\leq j} \frac {A^{ij} e^{-a^{ij} r_{ij}}} {r_{ij}}+ \sum_{i\leq j}B^{ij} e^{-b^{ij} r_{ij}^2 },
\end{eqnarray}
where the second term is a Gaussian which acts as a steep  repulsive wall and accounts for the anion hard core; these extra terms are used in cases where the ions are strongly polarized to avoid instability problems at very small anion-cation separations \cite{castiglione1999a}. The polarization part of the potential incorporates dipolar effects only. This reads:
\begin{eqnarray}
V^{\rm pol}&=&\sum_{i,j}-\left( q_i\mu_{j,\alpha} f_4^{ij}(r_{ij})- q_j\mu_{i,\alpha} f_4^{ji}(r_{ij})\right)T_{\alpha}^{(1)}({\bf r}_{ij}) \nonumber \\
\end{eqnarray}
\begin{equation} T_{\alpha}^{(1)}({\bf r})=-r_{\alpha}/r^3 \;\;\;\;\;\;\;\;\; T_{\alpha \beta}^{(2)}({\bf r})=(3r_{\alpha} r_{\beta}-r^2\delta_{\alpha \beta})/r^5. \end{equation}
The instantaneous values of these moments are obtained by  minimization of this expression with respect to the dipoles of all ions at each MD timestep. This  ensures that we regain the condition that the dipole  induced by an electrical field ${\bf E}$ is $\alpha {\bf E}$ and that the dipole values are mutually consistent. The short-range induction effects on the dipoles are taken into account by the Tang-Toennies damping functions:
\begin{equation}
 f_n^{ij}(r_{ij} )=1-c^{ij} e^{-b^{ij} r_{ij}} \sum_{k=0}^n \frac{(b^{ij}
 r_{ij})^k}{k!}.
\end{equation}
The parameters $b^{ij}$ determine the range at which the overlap of the charge densities affects the induced dipoles, the parameters $c^{ij}$ determine the strength of the ion response to this effect.}

\newpage

\bibliography{references}

\end{document}